\documentclass[conference]{IEEEtran}
\IEEEoverridecommandlockouts
\usepackage{cite}
\usepackage{amsmath,amssymb,amsfonts}
\usepackage{algorithm}
\usepackage{algpseudocode}
\usepackage{graphicx}
\usepackage{multicol}
\usepackage{multirow}
\usepackage{textcomp}
\usepackage{xcolor}
\usepackage{url}
\def\BibTeX{{\rm B\kern-.05em{\sc i\kern-.025em b}\kern-.08em
    T\kern-.1667em\lower.7ex\hbox{E}\kern-.125emX}}
\begin{document}

\title{Personalized and Adaptive Federated Learning for Canine Cardiomegaly Keypoint Prediction\\
}

\author{\IEEEauthorblockN{1\textsuperscript{st} Jialu Li}
\IEEEauthorblockA{
\textit{Yeshiva University}\\
New York, USA \\
jli10@mail.yu.edu}
\and
\IEEEauthorblockN{2\textsuperscript{nd} Youshan Zhang}
\IEEEauthorblockA{
\textit{Chuzhou University}\\
Chuzhou, China \\
youshan\_zhang@chzu.edu.cn}
\and
\IEEEauthorblockN{3\textsuperscript{rd} Honggang Wang}
\IEEEauthorblockA{
\textit{Yeshiva University}\\
New York, USA \\
honggang.wang@yu.edu}
}

\maketitle

\begin{abstract}
In the field of medical imaging, limited data and cross-institutional heterogeneity remain major challenges for training robust deep learning models. Federated learning (FL) has emerged as a promising approach that enables collaborative training across institutions while preserving data privacy by keeping datasets local. In this work, we propose a structure-aware personalized and adaptive federated learning framework, PAFL, for automatic vertebral heart score (VHS) estimation and cardiomegaly classification from canine thoracic X-rays. Our method treats the task as anatomical keypoint prediction followed by geometry-based VHS computation, introducing implicit structural consistency and improving interpretability. To address cross-institutional heterogeneity, we adopt a parameter-decomposed architecture that separates shared feature representations from client-specific adaptation modules. Furthermore, we introduce an adaptive, quality-aware aggregation strategy that dynamically weights client contributions based on both dataset size and model quality, thereby enhancing robustness to non-IID data distributions. Experimental results demonstrate that the proposed PAFL framework improves models' predictive performance compared to isolated training and to conventional and existing personalized federated learning (PFL) methods. Our results demonstrate the potential of integrating structure-aware modeling with PFL for reliable and interpretable medical image analysis.
\end{abstract}


\section{Introduction}
The rapid development of deep learning techniques has greatly advanced medical image analysis in both human and veterinary medicine. However, training high-performing deep learning models typically requires large-scale, diverse, and annotated datasets. In clinical environments, such datasets are often difficult to obtain due to privacy restrictions and limited annotations. Consequently, models trained on small or institution-specific datasets often suffer from limited generalization when deployed in new clinical settings. Federated learning (FL) has recently emerged as a promising framework for collaborative model training without requiring the cross-sharing of raw data. In an FL framework, each participating client independently trains an initial, centrally distributed model using its local data; only model weights are transmitted to a central server for aggregation~\cite{guan2024federated}. This decentralized training strategy preserves data privacy while supporting cross-institutional collaboration, enabling multiple institutions to benefit from larger and more diverse training data~\cite{nazir2023federated}. 

Recent advances in personalized federated learning attempt to address data heterogeneity by decomposing models into shared and client-specific components~\cite{collins2021exploiting, yang2023dynamic, sabah2024model, lai2025fedx}. While effective, these approaches assume the shared global representation generalize across clients without considering task-specific structural constraints. In medical imaging tasks, this assumption is often violated. In particular, directly applying FL methods to medical imaging analysis tasks may lead to two issues: (1) \textbf{structural inconsistency}, where models may fail to preserve geometric relationships between anatomical landmarks, and (2) \textbf{domain inconsistency}, where shared representations are biased toward dominant clients, reducing performance on underrepresented domains.

Among different canine diseases, cardiomegaly, defined as the enlargement of the cardiac silhouette~\cite{mourya2018vertebral}, is a critical indicator of potentially life-threatening cardiac conditions, calling the attention of both pet owners and the veterinary field. To standardize diagnostic procedures for cardiomegaly, the Vertebral Heart Score (VHS), proposed by Buchanan and Bücheler~\cite{buchanan1995vertebral}, has become a widely used clinical metric for assessing heart size in canine thoracic X-rays. Although VHS provides a standardized, reproducible measurement method, manual estimation is time-consuming and prone to variability, especially when comparing across breeds and imaging conditions. Automatic estimation of VHS from X-rays, therefore, is favored for both improved efficiency and diagnostic consistency~\cite{li2024regressive, mekonnen2025deep, nguemo2026automated}. However, the generalization of deep learning models in veterinary practices remains limited, and the FL training still faces several challenges. Data collected from different veterinary hospitals may show substantial heterogeneity due to variations in imaging equipment and patient populations. Such heterogeneity can reduce the effectiveness of conventional federated learning algorithms that rely on fully shared model parameters or naive aggregation.

To address these limitations, we propose a structure-aware PFL framework, PAFL, for automatic VHS estimation and cardiomegaly classification. Instead of directly predicting clinical scores, our method performs VHS estimation through anatomical keypoint detection followed by geometric computation. This design better preserves structural consistency and improves interpretability. Furthermore, we integrate this two-stage process into a PFL framework that separates shared feature representations from client-specific adaptation modules, enabling robust learning under heterogeneous data distributions. To further enhance robustness, we introduce a quality-aware aggregation strategy that dynamically weights client contributions based on both dataset size and model quality, mitigating the negative impact of low-quality or under-trained clients. The main contributions of this paper are as follows:
\begin{itemize}
    \item \textbf{Multi-institutional veterinary dataset.} We construct and utilize three canine thoracic X-ray datasets from institutions in the US, China, and Korea, capturing real-world heterogeneity in veterinary imaging. 
    \item \textbf{Structure-aware personalized federated learning.} We propose a federated learning framework that combines anatomical keypoint modeling with parameter decomposition between shared and client-specific, enabling robust learning under cross-institutional data heterogeneity, assisted by a domain adapter module in client models.
    \item \textbf{Quality-aware aggregation strategy.} We design an adaptive aggregation mechanism that accounts for both dataset size and model quality based on relative client performance, improving training stability and model generalization across heterogeneous settings.
\end{itemize}

\section{Methods}
\label{sec:methods}

\subsection{Overview}

In veterinary imaging, training well-performing models faces two primary challenges: structural inconsistency in clinical measurement tasks, which arises when models fail to preserve anatomical geometry, and domain heterogeneity across institutions, which stems from institutional variations. To address these challenges, our PAFL framework is designed with three objectives: (1) maintaining anatomical structures through keypoint modeling, (2) separating shared and personalized parameters to handle heterogeneity, and (3) adaptively weighting client contributions based on model quality.

\begin{figure}
    \centering
    \includegraphics[width=0.9\linewidth]{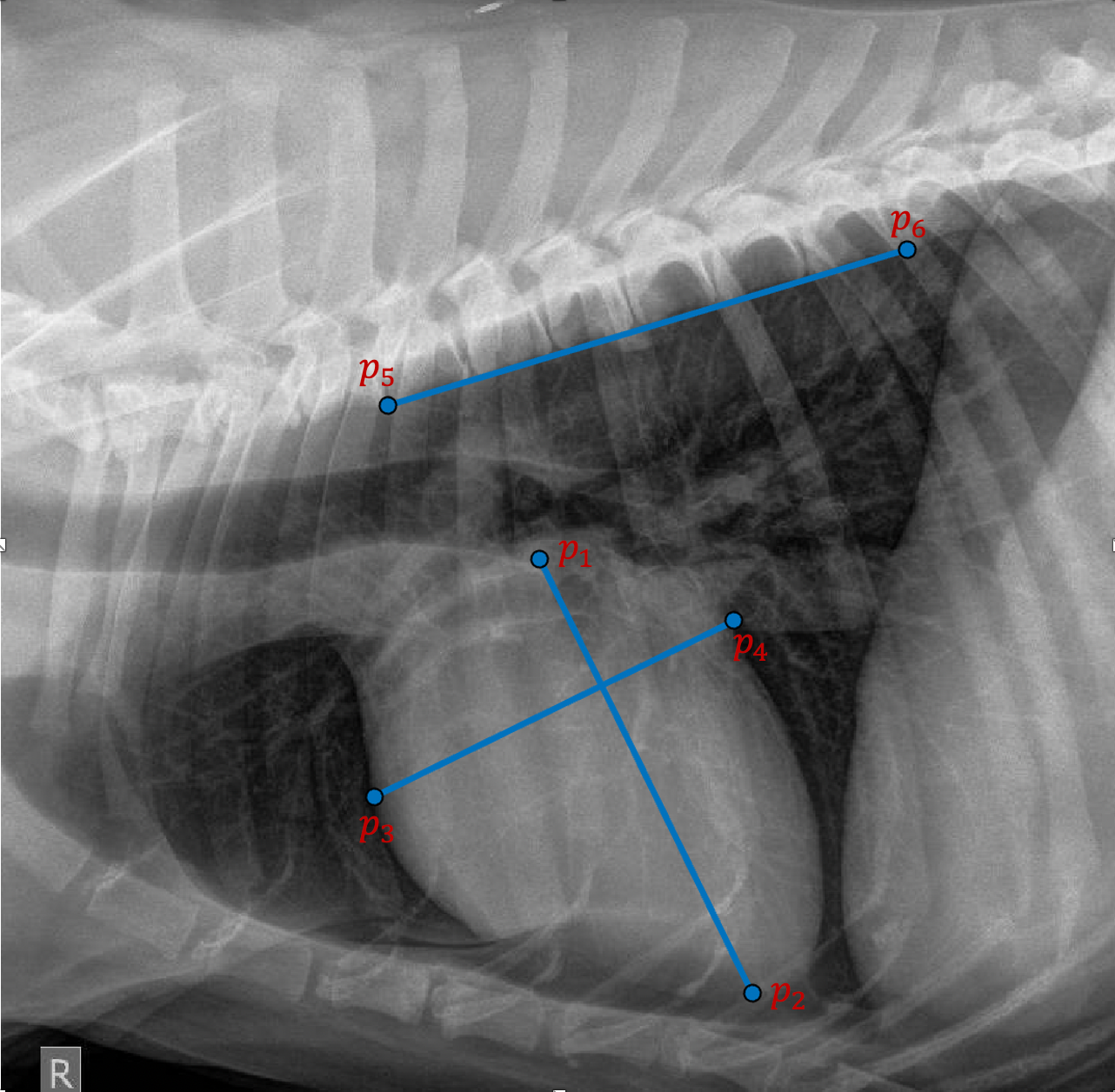}
    \caption{An annotated example of canine thoracic X-ray image.}
    \label{fig:x-ray}
\end{figure}


\subsection{Preliminaries}

We propose a federated learning framework with $N$ clients, where each client $k$ possesses a private dataset $\mathcal{D}_k = \{(x_i, y_i)\}_{i=1}^{n_k}$ following an underlying distribution $\mathcal{P}_k$, where $x_i \in \mathbb{R}^{H \times W \times 3}$ denotes a thoracic X-ray and $y_i \in \mathbb{R}^{12}$ represents six anatomical keypoints. Unlike standard supervised learning, the client distributions across datasets $\{\mathcal{P}_k\}_{k=1}^N$ are non-identical, reflecting domain heterogeneity across institutions. 

\paragraph{Personalized federated learning}
To address heterogeneity, PFL decomposes the model as:
\begin{equation}
\theta_k = \{\theta_b, \theta_{k}\},
\end{equation}
where $\theta_b$ is shared and $\theta_{k}$ is client-specific.

Our objective is to train a set of client-specific models $\{f_k\}$ and minimize the average errors across clients while accounting for the heterogeneity of local data distributions:

\begin{equation}
\min_{\{\theta_k\}} \sum_{k=1}^N \pi_k \, \mathcal{L}_k(\theta_k),
\end{equation}
where $\pi_k = \frac{n_k}{\sum_j n_j}$ denotes the relative dataset size, and the local objective is defined as:
\begin{equation}
\mathcal{L}_k(\theta_k) = \frac{1}{|\mathcal{D}_k|}
\sum_{i=1}^{|\mathcal{D}_k|}
\left\| f_{\theta_k}(x_i) - y_i \right\|_2^2,
\end{equation}
where $f_{\theta_k}$ denotes the client-specific model by $\theta_k$, and $\mathcal{D}_k$ is the local dataset of client $k$.

\paragraph{Keypoint-based clinical metric measurement} In canine cardiomegaly assessment, the VHS is computed from six anatomical keypoints. The calculation sums the lengths of the long and short cardiac axes and then normalizes them to the vertebral column, starting at the T4 vertebral level~\cite{buchanan1995vertebral}. An example of canine thoracic X-ray is presented in Fig~\ref{fig:x-ray} with six keypoints annotated. Given six keypoints $\{p_i\}_{i=1}^6$, the geometric calculation for VHS is defined as:
\begin{equation}
\text{VHS} = \frac{6\left( \|p_1 - p_2\|_2 + \|p_3 - p_4\|_2 \right)}{\|p_5 - p_6\|_2}.
\end{equation}


\subsection{Structure-Aware Anatomical Learning via Keypoint prediction}

\paragraph{Motivations}
Existing approaches treat VHS estimation as a direct regression or classification problem. They typically approximate a mapping $f: x \rightarrow v$ for VHS estimation. However, such mappings ignore the anatomical structure when measuring clinical metrics, leading to lack of interpretability in predictions, undermining the model's trustworthiness. 

\paragraph{Structure-aware prediction}
To overcome this limitation, we treat VHS estimation as a structured geometric prediction problem. The prediction process is expressed as:
\begin{equation}
x \xrightarrow{f_\theta} \hat{y} \xrightarrow{\mathcal{G}} \hat{v},
\end{equation}
where $\hat{y} \in \mathbb{R}^{12}$ denotes predicted keypoint coordinates and $\mathcal{G}(\cdot)$ is a deterministic geometric formula.
Let $\hat{y} = \{p_1,\dots,p_6\}$ with $p_i \in \mathbb{R}^2$. We define:
\begin{equation}
d_1 = \|p_1 - p_2\|_2, \quad
d_2 = \|p_3 - p_4\|_2, \quad
d_3 = \|p_5 - p_6\|_2,
\end{equation}
and
\begin{equation}
\hat{v} = \mathcal{G}(\hat{y}) = \frac{6(d_1 + d_2)}{d_3}.
\end{equation}

The classification head, a hard-coded classification algorithm, then classifies the cardiomegaly condition in X-rays through $\hat{v}$ using the predefined threshold: :

\begin{equation}
\hat{c} =
\begin{cases}
0, & \text{if } \hat{v} < 8.2,\\
1, & \text{if } 8.2 \leq \hat{v} \leq 10,\\
2, & \text{if } \hat{v} > 10.
\end{cases}
\label{eq:13}
\end{equation}

Incorporating the clinical keypoint prediction and measurement process into the model training ensures that models learn spatially consistent, anatomically meaningful keypoints, minimizing geometric error. 

\begin{figure*}
    \centering
    \includegraphics[width=0.9\linewidth]{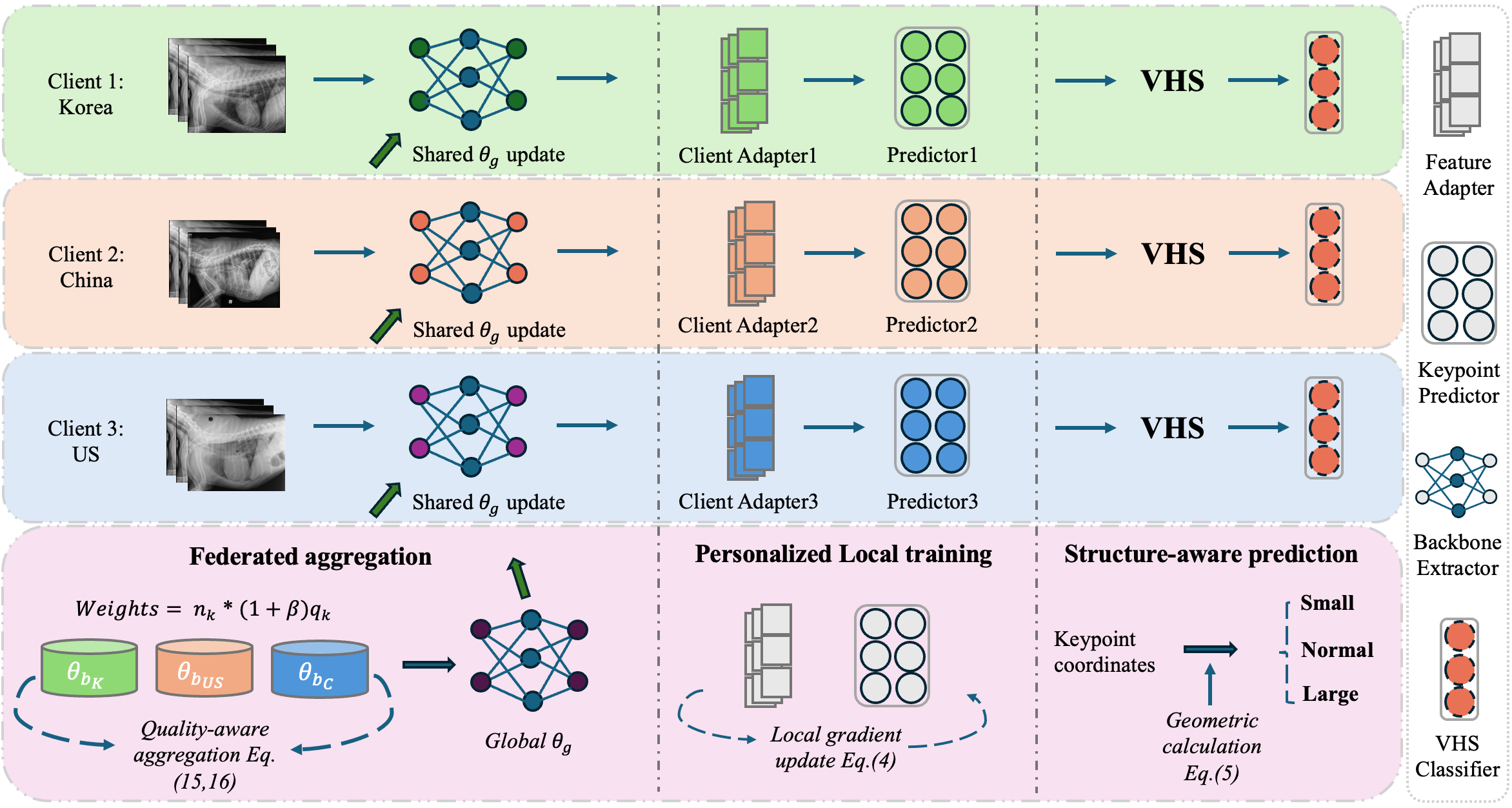}
    \caption{Proposed personalized and adaptive federated learning architecture for quality-aware aggregation.}
    \label{fig:fed_model}
\end{figure*}

\subsection{Personalized Federated Learning under Heterogeneity}

\paragraph{Motivation}
Federated learning aims to leverage distributed data across clients to train more generalizable models, but standard approaches assume a shared representation across all clients. In practice, veterinary datasets demonstrate significant heterogeneity, making a fully shared model insufficient to improve predictive performance. At the same time, fully local models fail to benefit from cross-institutional knowledge.

Therefore, in our proposed framework, we design and decompose each client model into a shared backbone, a personalized adapter, and a prediction head:
\begin{equation}
f_\theta(x) = h_{\theta_h} \big( a_{\theta_a} ( b_{\theta_b}(x) ) \big),
\end{equation}
where $\theta = \{\theta_b, \theta_a, \theta_h\}$, $b$, $a$, and $h$ are for backbone, adapter, and regression head.

\paragraph{Shared representation}
The backbone $b_{\theta_b}$ extracts latent representations $z \in \mathbb{R}^d$ from input X-rays:
\begin{equation}
z = b_{\theta_b}(x),
\end{equation}
capturing domain-invariant anatomical features.

\paragraph{Adaptive transformation}
The adapter is composed of a 2-layer MLP with a ReLU activation. It acts as a residual feature transformation:
\begin{equation}
a_{\theta_a}(z) = z + \phi_{\theta_a}(z),
\end{equation}
where $\phi_{\theta_a}$ is a learnable nonlinear mapping. The incorporation of this adapter module ensures that features extracted using the backbone with globally shared weights can be dynamically adapted to the local data distributions.

\paragraph{Personalized prediction}
The prediction head outputs normalized keypoint coordinates:
\begin{equation}
\hat{y} = h_{\theta_h}(a_{\theta_a}(z)).
\end{equation}

The backbone captures universal anatomical features that are consistent across institutions, such as cardiac shape and spatial layout. The adapter and prediction head allow each client to adjust to local variations, including breed distribution and imaging conditions. This parameter separation prevents conflicting optimization objectives between global and local knowledge, mitigating learning compromise across heterogeneous clients. 

\paragraph{Personalized federated optimization}

We adopt a personalized parameter sharing strategy:
\begin{equation}
\theta_k = \{\theta_b, \theta_{a,k}, \theta_{h,k}\},
\end{equation}
where $\theta_b$ is shared and $\{\theta_{a,k}, \theta_{h,k}\}$ are client-specific.

Each client trains following the minimization of the mean squared error (MSE) objective:

\begin{equation}
\mathcal{L}_k = \frac{1}{|\mathcal{D}_k|} 
\sum_{i=1}^{|\mathcal{D}_k|} 
\left\| f_\theta(x_i) - y_i \right\|^2.
\end{equation}

During this process, each local model learns a shared feature extractor with task-specific heads. After training, only the shared backbone is transmitted to the server, preserving privacy while enabling collaboration.

\subsection{Quality-Aware Federated Aggregation}

\paragraph{Motivation}
A critical limitation of standard FedAvg is that it treats all clients equally or relies solely on dataset size. However, in heterogeneous clinical environments, clients may differ not only in data quantity but also in data and model quality. Aggregating low-quality or under-trained models can degrade global performance. 

\paragraph{Quality-aware weighting}
In our proposed framework, the classification accuracy was calculated as a geometric derivative using threshold-based metrics, making it unsuitable for capturing fine-grained differences and thus for accurately assessing the model's performance in keypoint coordinate prediction. Therefore, compared with other works~\cite{deng2022improving,pejo2023quality,peng2025federated}, we select validation loss, derived using the EfficientNet-B7 backbone~\cite{tan2019efficientnet}, as the quality metric as it provides a more stable measure of model performance on keypoint prediction.

Instead of directly using the absolute validation loss per client, which may not be comparable due to heterogeneity, we define a normalized relative quality score:
\begin{equation}
q_k = \frac{\max_j \mathcal{L}_j^{\text{val}} - \mathcal{L}_k^{\text{val}}}
{\max_j \mathcal{L}_j^{\text{val}} - \min_j \mathcal{L}_j^{\text{val}} + \epsilon},
\label{eq:15}
\end{equation}
where $j \in  {1,2,...,N}$ for clients, and $\epsilon$ is a small constant to ensure numerical stability, in our case, $1\times{10}^{-12}$.

\paragraph{Weighted Aggregation}
To balance the dataset size and model quality during our weight aggregation process, we define the aggregation weight as:
\begin{equation}
w_k = \frac{n_k \left(1 + \beta q_k \right)}
{\sum_{j=1}^N n_j \left(1 + \beta q_j \right)},
\label{eq:16}
\end{equation}
where $\beta$ controls the influence of the quality term. We set $\beta=0.3$ to avoid over-amplifying noisy or biased clients under heterogeneous data distributions.
 
The global backbone update is then computed as:
\begin{equation}
\theta_b^{(t+1)} = \sum_{k=1}^N w_k \theta_{b,k}^{(t)}.
\label{eq:17}
\end{equation}

By normalizing validation losses across clients, our proposed framework ensures that model quality is comparable across clients with non-IID distributions. The term $(1 + \beta q_k)$ allows higher-quality clients to contribute more while preventing over-amplification for larger datasets. The comprehensive federated learning architecture is shown in Fig.~\ref{fig:fed_model}.

\subsection{Model reconstruction and personalization}

After aggregation, each client receives the global weight and reconstructs its model:
\begin{equation}
\theta_k^{(t+1)} = \{\theta_b^{(t+1)}, \theta_{a,k}^{(t)}, \theta_{h,k}^{(t)}\}.
\end{equation}

This design enables the model to retain globally shared anatomical features while adapting to local domain characteristics, improving both generalization and robustness towards domain shift and client heterogeneity. 

\paragraph{Early stopping} Our framework employs a client-aware early stopping mechanism. We monitor validation loss for each client individually and define improvement as a reduction in loss of more than $1 \times 10^{-6}$ for any client. The training process continues as long as at least one client shows progress, and terminates when no client improves for 10 consecutive rounds. 

\section{Experiments}
\label{sec:exp}

\subsection{Datasets}
In this paper, we utilize canine thoracic X-ray images collected from three countries: 2,000 from China~\cite{li2024regressive}, 1,600 from the United States, and 1,000 from the Republic of Korea. The datasets from China and the US were collected manually by the research group members, and the VHS ground-truth keypoint labels for all three datasets were generated using software developed with a few-shot generalization strategy, validated by professional vets. The China and Korea datasets follow a standard 70/10/20 training/validation/test split, while the U.S. dataset uses 700 training data points due to the data availability. The detailed dataset statistics are displayed in Tab.~\ref{tab:dataset}. The last column indicates the class distributions are relatively balanced across three datasets.

\begin{table}[t]
\caption{Statistics for three client datasets}
\label{tab:dataset}
\begin{center}
\setlength{\tabcolsep}{+2mm}{
\begin{tabular}{ccccc}
\hline
&Train & Valid & Test & Cardiomegaly pct(\%)\\
\hline
China & 1,400 & 200 & 400 & 870/43.5\%\\
United States & 700 & 300 & 600 & 655/40.9\%\\
Korea & 700 & 100 & 200 & 568/56.8\%\\
\hline
\end{tabular}}
\end{center}
\end{table}

\begin{table*}[t]
\caption{Federated learning results on local datasets with 30 rounds of global training.}
\label{tab:results}
\begin{center}
\setlength{\tabcolsep}{+5mm}{
\begin{tabular}{cccccccc}
\hline
 \multirow{2}{*}{Federated Learning}
& \multicolumn{2}{c}{Server 1 (China)} & \multicolumn{2}{c}{Server 2 (US)} & \multicolumn{2}{c}{Server 3 (Korea)} \\
\cline{2-7}
& Validation & Test  & Validation & Test & Validation & Test  \\
\hline
Round 1 (initial) & 77.5\% & 76.0\% & 72.0\% & 71.0\% & 74.3\% & 76.0\% \\
Round 5 & 78.5\% & 78.0\% & 79.7\% & 80.6\% & 81.0\% & 82.0\% \\
Round 10 & 82.5\% & 82.5\% & 78.6\% & 82.2\% & 81.0\% & 87.0\% \\
Round 15 & 82.0\% & 85.5\% & 82.7\% & 83.5\% & 82.5\% & 85.5\%  \\
Round 20 & 81.3\% & 86.5\% & 80.6\% & 84.5\% & 82.0\% & 88.0\%  \\
\hline
\textbf{Round 25} & 83.6\% & \textbf{88.5}\% & 83.0\% & \textbf{86.0}\% & 82.5\% & \textbf{89.5\%}  \\
\hline
Round 30 & 81.0\% & 88.0\% & 81.0\% & 85.2\% & 84.0\% & 89.0\%  \\

\hline
\end{tabular}}
\end{center}
\end{table*}

\subsection{Implementation details}

For the federated learning framework, we performed 30 rounds of global communication.
Each local server trained its model on its own dataset for 50 epochs before sending the updated weights to the subsequent central aggregation process. The central server performed the adaptive weight aggregation based on dataset sizes and model training quality. The framework was run on three A100 GPUs in parallel with an image resolution of 512 $\times$ 512, a default batch size of 8, a learning rate of $3 \times 10^{-5}$, and using an Adam optimizer. We select the EfficientNet-B7 as the backbone network for our experiments. Source code, training algorithm and more experiments are available at \url{https://github.com/YSZ-by/PAFL}.

\subsection{Results}

\begin{table*}[t]
\caption{Federated learning methods comparison on local datasets with  10 rounds of global training.}
\label{tab:compare}
\begin{center}
\setlength{\tabcolsep}{+5.5mm}{
\begin{tabular}{cccccccc}
\hline
 \multirow{2}{*}{Methods}
& \multicolumn{2}{c}{Server 1 (China)} & \multicolumn{2}{c}{Server 2 (US)} & \multicolumn{2}{c}{Server 3 (Korea)} \\
\cline{2-7}
& Validation & Test  & Validation & Test & Validation & Test  \\
\hline
FedProx~\cite{li2020federated} & 78.0\% & 79.0\% & 76.0\% & 79.8\% & 81.0\% & 83.5\% \\
FedCross~\cite{hu2024fedcross} & 81.2\% & 81.8\% & 80.0\% & 80.8\% & 81.5\% & 83.0\% \\
RobustPFL~\cite{chen2025robustpfl} & 82.5\% & 81.0\% & 81.0\% & 80.0\% & 84.0\% & 84.5\% \\
PSFL~\cite{dai2025psfl} & 83.5\% & \textbf{82.5\%} & 81.0\% & 81.5\% & 82.0\% & 86.5\% \\
\hline

\textbf{PAFL}  & 82.5\% & \textbf{82.5\%} & 78.6\% & \textbf{82.2\%} & 81.0\% & \textbf{87.0\%} \\

\hline
\end{tabular}}
\end{center}
\end{table*}

\begin{table*}[t]
\caption{Ablation study for the key components of the PAFL framework with 12 rounds of global training.}
\label{tab:ablation}
\begin{center}
\setlength{\tabcolsep}{+1.5mm}{
\begin{tabular}{cccccccccc}
\hline
 \multirow{2}{*}{Model} & \multirow{2}{*}{Personalization}
&  \multirow{2}{*}{Quality-aware} & \multirow{2}{*}{Relative quality} & \multicolumn{2}{c}{Server 1 (China)} & \multicolumn{2}{c}{Server 2 (US)} & \multicolumn{2}{c}{Server 3 (Korea)} \\
\cline{5-10}
& & & & Validation & Test  & Validation & Test & Validation & Test  \\
\hline
1&$\times$ & $\times$ & $\times$ & 81.5\% & 82.3\% & 79.3\% & 81.7\% & 81.6\% & 82.5\% \\
2&$\checkmark$ & $\times$ & $\times$ & 81.5\% & \textbf{85.8\%} & 81.0\% & 82.0
\% & 82.0\% & 86.5\% \\
3&$\times$ & $\checkmark$ & $\checkmark$ & 80.5\% & 82.5\% & 80.0\% & 81.2\% & 83.0\% & 85.0\%\\
4&$\checkmark$ &$\checkmark$ & $\times$ & 76.5\% & 81.8\% & 73.3\% & 70.0\% & 77.0\% & 80.0\% \\
\hline
5&$\checkmark$ &$\checkmark$ & $\checkmark$ & 82.5\% & 84.25\% & 81.3\% & \textbf{84.5\%} & 83.0\% & \textbf{88.5\%} \\
\hline
\end{tabular}}
\end{center}
\end{table*}

For federated learning, we trained the entire PAFL framework for 30 consecutive rounds, limited by training time. The central server received the model weights from individual clients and performed the adaptive weight aggregation using methods described in Sec.~\ref{sec:methods}. The results for the PAFL training are displayed in Tab.~\ref{tab:results}. The results demonstrated that the model learned actively during the first 20 rounds, with every 5 steps showing accuracy improvements, then became relatively stagnant and fluctuated, but still showed reasonable improvement in subsequent training steps. Different clients exhibit different improvement rates, while we see that all client models performed best primarily around their 25th training step, achieving optimal performance in keypoint prediction and VHS calculation on local test sets, yielding the highest accuracy scores. Compared to the initial round of training, the accuracy scores in the 25th round improved by more than 10\%, demonstrating the effectiveness of our personalized and adaptive federated learning framework in enhancing the deep learning model's performance compared to isolated training. 


Additionally, we compare our PAFL framework with existing federated learning methods, traditional FL methods FedProx~\cite{li2020federated} and FedCross~\cite{hu2024fedcross}, as well as personalized FL frameworks, RobustPFL~\cite{chen2025robustpfl}, and PSFL~\cite{dai2025psfl}. The comparison models were mostly reproduced from the method sections of the original papers or adapted from the source code for FedProx~\cite{FedProx}. All models were trained for 10 rounds and compared to the performance at the 10th round of the proposed PAFL framework. The results in Tab.~\ref{tab:compare} show that our model outperformed traditional FL frameworks and RobustPFL, and achieved performance comparable to or better than PSFL, indicating the promising applicability of our PAFL framework to improved medical imaging analysis tasks.

\subsection{Discussion}
From the above results, we can see that our proposed PAFL framework improves deep learning model performance by expanding data availability and diversity for training while ensuring the model's resistance to dataset heterogeneity. The results demonstrated at least a 10\% increase in prediction performance across all local client servers, indicating promising aggregation outcomes.

To better analyze the efficacy of each module in our proposed PAFL framework, we conduct ablation studies to evaluate the performance of our model's personalization strategy and quality-aware aggregation method. The results displayed in Tab.~\ref{tab:ablation} are the classification accuracy for 12 rounds of federated training. We conducted five studies, including comparisons of raw and relative quality computation, of personalized and non-personalized model aggregation, and of traditional FedAvg and quality-aware FedAvg. From Tab.~\ref{tab:ablation}, we can see that the choice of quality metric matters greatly for model performance, comparing model 4 (using individual raw MSE loss) and model 5 (using relative MSE loss across clients). Poor quality evaluation degraded overall model performance. Additionally, personalized models that aggregated only backbone parameters while preserving local feature adapters yielded better results than models fully aggregated via FL frameworks. 

Additionally, during PAFL training, we found that frameworks without quality-aware aggregation always converged prematurely, around the 10th round, resulting in limited improvement in later stages compared to quality-aware FL. This finding explains why the PAFL framework achieved comparable or slightly better performance in the first dozen rounds (Tab.~\ref{tab:ablation}) and surpassed other frameworks as training rounds increased, as shown in Tab.~\ref{tab:results}. This phenomenon demonstrates the promising application of our proposed PAFL framework in increasing model prediction accuracy while being more resistant to premature convergence.

\section{Conclusions}

In this paper, we present a structure-aware personalized and adaptive federated learning framework (PAFL) for automatic VHS estimation from canine thoracic X-rays. 
We applied a personalized federated learning strategy that decomposes the model into a shared backbone and client-specific adaptation and prediction modules. Moreover, we introduced a quality-aware aggregation mechanism that leverages relative validation performance to dynamically adjust client contributions, improving the FL framework's robustness under non-IID data distributions. Extensive experiments showed that our proposed PAFL framework achieved state-of-the-art performance.
The results demonstrated the promise of applying a personalized and quality-aware FL framework to medical imaging tasks to address training data scarcity and heterogeneity.


\bibliographystyle{plain}
\bibliography{ref}


\end{document}